# Weaving, Bending, Patching, Mending the Fabric of Reality: A Cognitive Science Perspective on Worldview Inconsistency

Liane Gabora

**ABSTRACT**

In order to become aware of *inconsistencies*, one must first construe of the world in a way that reflects its *consistencies*. This paper begins with a tentative model for how a set of discrete memories transforms into an interconnected worldview wherein relationships between memories are forged by way of abstractions. Inconsistencies prompt the invention of new abstractions. In regions of the conceptual network where inconsistencies abound, a cognitive analog of simulated annealing is in order; there is a willingness to question previous assumptions—to 'loosen' conceptual relationships—so as to let new concepts thoroughly percolate through the worldview and exert the needed revolutionary effect. In so doing there is a risk of assimilating dangerous concepts. Repression arrests the process by which dangerous thoughts infiltrate the conceptual network, and deception blocks thoughts that have already been assimilated. These forms of self-initiated worldview inconsistency may evoke feelings of fragmentation at the level of the individual or the society.

## 1. INTRODUCTION

This volume addresses the question of how we detect and cope with worldview inconsistencies. The present paper will explore how inconsistencies promote abstractions that resolve them, and that sometimes percolate through the worldview and reconfigure it. But perhaps it is appropriate to begin by backing up to consider a more basic question: how is it that the human mind is aware of *consistencies* in its world? This is, in fact, no small feat. The primate mind, for example, is incapable of linking concepts together in a chain of associations and thereby determining the degree to which they are consistent or inconsistent; it makes no attempt to stitch together strands of experience to create a coherent internal model of the fabric of reality. There is much evidence that this proclivity came into existence with the arrival of *Homo erectus* approximately 1.7 million years ago (Donald 1991). Section Two outlines a model of how this cognitive transition may have occurred. Since the model is presented in detail elsewhere (Gabora, 1998) it is explained here only insofar as is necessary to make sense of the material that follows.

Armed with an idea of how we forge and incrementally elaborate a more or less consistent internal model of the world, we are in a better position to understand what happens when inconsistencies are encountered. Section Three discusses how we 'patch holes' in our worldview by inventing abstractions, and how formidable inconsistencies prompt us to 'unravel and reweave', a process that can be modeled using a cognitive analog of simulated annealing.

Section Four deals with worldview inconsistencies that we introduce voluntarily (albeit often unconsciously) through censorship, repression, and deception. Censorship artificially curtails the process by which a new experience or idea gets woven into the individual's conceptual web. Deception perpetuates more deception, much as a fold tends to perpetuate itself along a length of fabric. These processes are less like to manifest as logical paradoxes than as generalized feelings of fragmentation.

Section Five takes a closer look at the 'worldview as tapestry' metaphor that has been developed in the previous sections, and attempts to tie the paper together, so to speak, by clarifying this metaphor.

The paper is speculative, heavier on theory than data (to say the least), so it won't be everyone's cup of tea. My goal in writing it was to express some provocative, interdisciplinary ideas that would round out this volume with a cognitive science perspective.

## 2. WEAVING DISCRETE MEMORIES INTO A COHERENT WORLDVIEW

### 2.1 A Transition in Cognitive Capacity

The early human memory system appears to have been, like that of a primate, limited to the storage and cued retrieval of specific episodes (see Heyes, 1998). Accordingly, Donald (1991) uses the term *episodic* to designate a mind such as this that consists only of episodic memories, no abstractions. The awareness of an episodic mind is dominated by the events of the present moment. Occasionally it encounters a stimulus that is similar enough to some stored episode to evoke a retrieval or reminding event, and sometimes the stimulus evokes a reflexive, or (with much training) learned response. However, it has great difficulty accessing memories independent of environmental cues. It can not manipulate symbols and abstractions, or invent them on its own, and is unable to improve skills through self-cued rehearsal. It seems to encode each item as a separate, self-contained, unmodifiable entity.

In contrast, the modern human mind encodes relationships between episodes by way of abstractions, and relates abstractions to one another by way of higher-order abstractions. For example, we know that the experience of seeing Rover was similar to the experience of seeing Lassie because Rover and Lassie are both instances of 'dog', and we know that dogs are animals. The human mind can retrieve and recursively operate on memories independent of environmental cues, a process referred to by Karmiloff-Smith (1992) as *representational redescription*. By redescribing an episode in terms of what is already known, it gets rooted in the network of understandings that comprise the worldview, and in turn, the worldview is perpetually revised as new experiences are assimilated and new symbols and abstract concepts invented as needed. The capacity for a self-sustained stream of thought that both structures and is structured by an internal model of the world enables us to plan and predict, to generate novelty, to tailor behavior according to context, and to exert precise control over intentional communication.

The existence of this uniquely human form of cognition leaves us with a nontrivial question of origins. What sort of functional reorganization would turn an episodic mind into that of a modern human? In the absence of representational redescription, how are relationships established such that the memory turns into a vast, interwoven conceptual web? And until a memory incorporates relationships, how can one idea evoke another, which evokes another, *et cetera*, in a stream of representational redescription? In other words, if you need a worldview to generate a stream of thought, and streams of thought are necessary to connect knowledge into a worldview, how could one have come into existence without the other? We have a chicken-and-egg problem.

The explanation proposed here was inspired by an idea put forward to explain the origin of life. The origin of life and the origin of the cognitive dynamics underlying culture might appear at first glance to be very different problems. However, deep down, they both amount to the same thing: the bootstrapping of a system by which information patterns are generated, and the selective proliferation of some variants of these patterns over others (Dawkins 1975; Gabora 1997). Thus culture, like biology, can be viewed as a form of evolution, altbeit one that manifests very differently from biological evolution. Imitation and learning constitute the *replication phase* of this form of evolution, and creativity is the *variation-generating phase*. In keeping with this evolutionary framework, the term 'meme' is used to refer to a unit of cultural information as it is represented in the brain. Thus we take a broad interpretation of the meme concept. A meme can be anything from an idea for a recipe, to a memory of one's uncle, to a concept of size, to an attitude of racial prejudice. Episodic memories and symbolic abstractions are memes, as is any element rational or irrational, conscious or unconscious, that constitutes a component of a worldview. The rationale for lumping together these diverse elements is that they are all 'food for thought', units of information that can be drawn upon to invent new memes or to clarify relationships amongst existing ones. Memes that have been implemented as actions, vocalizations, or objects are referred to as artifacts. The memes that dwell in the mind of an individual 'host' work together to build the illusion of a cohesive worldview and unified ego, yet compete for attention to get expressed as words, actions, or objects in the physical world.

## 2.2 The Autocatalytic Theory of the Origin of Life

We will put the aside the origin of worldviews and culture for now, and turn to the question of biological origins. The origin of life paradox can be stated simply: if living things come into existence when other living things give birth to them, how did the first living thing arise? That is, how did something complex enough to reproduce itself come to be? In biology, self-replication is orchestrated through an intricate network of interactions between DNA, RNA, and proteins. DNA is the genetic code; it contains instructions for how to construct various proteins. Proteins, in turn, both catalyze reactions that orchestrate the decoding of DNA by RNA, and are used to construct a body to house and protect all this self-replication machinery. Once again, we have a chicken-and-egg problem. If proteins are made by decoding DNA, and DNA requires the catalytic action of proteins to be decoded, which came first? How could a system composed of complex, mutually dependent parts come into existence?

Kauffman (1991) suggested that once you have some sort of self-replicating structure in place, anything whatsoever that accomplishes this basic feat, natural selection can enter the picture and help things along. Accordingly, he decided to focus on how to get from no life at all to any kind of primitive self-replicating system, and hand the problem of getting from there to DNA-based life, over to natural selection (as well as self-organizing processes). Given the conditions present on earth at the time life began, how might some sort of self-replicating system have arisen? His answer is that life may have begun not with a single molecule capable of replicating *itself*, but with a set of *collectively* self-replicating molecules. That is, none of the molecules could replicate itself, but each molecule could induce the replication of some other molecule in the set, and likewise, its own replication was induced by some other member of the set. This kind of dual role as both ingredient (or stimulant) and product of different chemical reactions is not uncommon for polymers such as protein and RNA molecules.

Polymers induce each other's replication by acting as catalysts. Catalysts speed up chemical reactions that would otherwise occur very slowly. An autocatalytic system is a set of molecules which, as a group, catalyze their own replication. Thus if A catalyzes the conversion of X to B, and B catalyzes the conversion of Y to A, then A + B comprise an autocatalytic set (FIGURE 1). In an environment rich in X and Y, A + B can self-replicate. A set of polymers wherein each molecule's formation is catalyzed by some other molecule is said to exhibit catalytic closure.

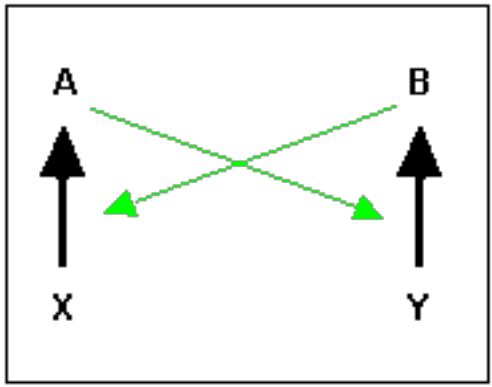

FIGURE 1. An autocatalytic set: A catalyses the formation of B, and B catalyses the formation of A. Thick black arrows represent catalyzed reactions. Thin green arrows represent catalysis.

It is of course highly unlikely that two polymers A and B that just happened to bump into one another would happen to catalyze each other. However, this is more likely than the existence of a single polymer catalyzing its own replication. And in fact, when polymers interact, their diversity increases, and so does the probability that some subset of the total reaches a critical point where there is a catalytic pathway to every member. To show that this is true we must show that $R$, the number of reactions by which they can interconvert increases faster than $N$, their total number. So long as this is true, then if each reaction has some probability of getting carried out, the system eventually undergoes a sharp transition to a state where there is a catalytic pathway to each polymer present. We find that $R/N$ does indeed increase. For the mathematical details, refer to (Kauffman 1993) or my summary in (Gabora, 1998).

This kind of sharp phase transition is a statistical property of random graphs and related systems such as this one. Random graphs consist of dots, or 'nodes', connected to each other by lines or 'edges'. As the ratio of edges to nodes increases, the probability that any one node is part of a chain of connected nodes increases, and chains of connected nodes become longer. When this ratio reaches approximately 0.5, almost all these short segments become cross-connected to form one giant cluster (FIGURE 4). Plotting the size of the largest cluster versus the ratio of edges to nodes yields a sigmoidal curve. The larger the number of nodes, the steeper the vertical portion of this curve (referred to as the *percolation threshold*).

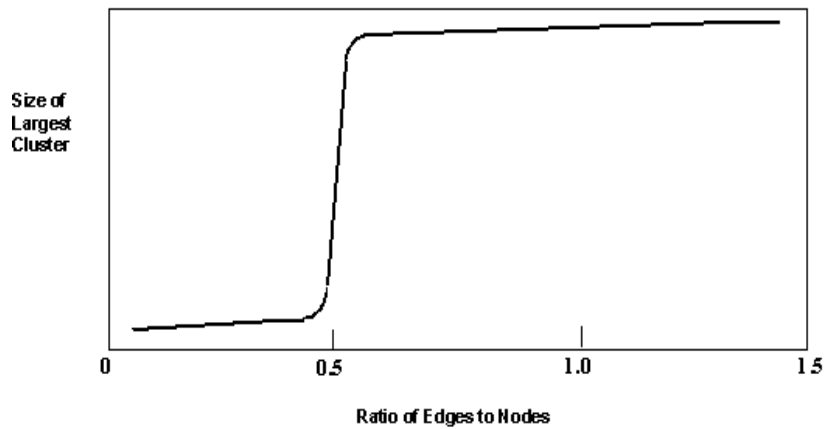

FIGURE 2. When the ratio of edges to nodes reaches approximately 0.5, short segments of connected nodes join to form a large cluster that encompasses the vast majority of nodes.

Of course, even if catalytic closure is theoretically possible, we are still a long way from knowing that it is the correct explanation for the origin of life. How likely is it that an autocatalytic set would have emerged given the particular concentrations of chemicals and atmospheric conditions present at the time life began? In particular, some subset of the $R$ theoretically possible reactions may be physically impossible; how can we be sure that every step in the synthesis of each member of an autocatalytic set will actually get catalyzed? Kauffman's response is: if we can show that autocatalytic sets emerge for a wide range of hypothetical chemistries—i.e., different collections of catalytic molecules—then the particular details of the chemistry that produced life do not matter so long as it falls within this range. We begin by noting that, much as several different keys sometimes open the same door, each reaction can be catalyzed by, not a single catalyst but a hypersphere of catalytic molecules, with varying degrees of efficiency. So we assign each polymer an extremely low *a priori* random probability $P$ of catalyzing each reaction. The lower the value of $P$, the greater $M$ must be, and vice versa. Kauffman shows that the values for $M$ and $P$ necessary to achieve catalytic closure with a probability of $> 0.999$ are highly plausible given the conditions of early earth. Experimental evidence for this theory using real chemistries (Lee *et al*. 1996, 1997; Severin *et al*. 1997), and computer simulations (Farmer *et al*. 1986) have been unequivocally supportive. Farmer *et al*. showed that in an 'artificial soup' of information strings capable of cleavage and ligation reactions, autocatalytic sets do indeed arise for a wide range of values of $M$ and $P$. FIGURE 3 shows an example of one of the simplest autocatalytic sets it produced. The original polymers from which an autocatalytic set emerges is referred to as the 'food set'. In this case it consists of 0, 00, 1, and 11. As it happens, the autocatalytic set that eventually emerges contains all members of the original food set. This isn't always the case.

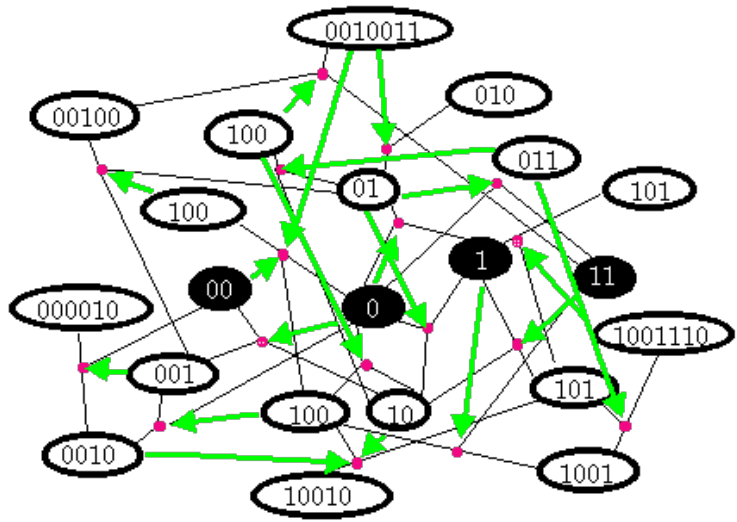

Figure 3. A typical example of a small autocatalytic set. Reactions are represented by thin, black lines connecting ligated polymers to their cleavage products. Thick, green lines indicate catalysts. Dark ovals represent food set.

An interesting question explored in this simulation is: once a set of polymers has achieved autocatalytic closure, does that set remain fixed, or is it able to incorporate new polymer species? They found that some sets were *subcritical*—unable to incorporate new polymers—and others were *supracritical*—incorporated new polymers with each round of replication. Which of these two regimes a particular set fell into depended on $P$, and the maximum length of the food set polymers.

Now the question is: supposing an autocatalytic set did emerge, how would it evolve? The answer is fairly straightforward. It is commonly believed that the primitive self-replicating system was enclosed in a small volume (such as a coascervate or liposome) to permit the necessary concentration of reactions (Oparin 1971; Morowitz 1992; Cemin & Smolin, in press). Since each molecule is getting duplicated somewhere in the set, eventually multiple copies of all molecules exist. The abundance of new molecules exerts pressure on the vesicle walls. This often causes such vesicles to engage in a process called *budding*, where it pinches off and divides into two 'twins'. So long as each twin contains at least one copy of each kind of molecule, the set can continue to self-replicate indefinitely. Replication is far from perfect, so an 'offspring' is unlikely to be identical to its 'parent'. Different chance encounters of molecules, or differences in their relative concentrations, or the arrival of new 'food' molecules, could all result in different catalysts catalyzing a given reaction, which in turn alters the set of reactions to be catalyzed. So there is plenty of room for heritable variation. Selective pressure is provided by the affordances and limitations of the environment. For example, say an autocatalytic set of RNA-like polymers arose. Some of its offspring might have a tendency to attach small molecules such as amino acids (the building blocks from which proteins are made) to their surfaces. Some of these attachments inhibit replication, and are selected against, while others favor it, and are selected for. We now have the beginnings of the kind of genotype-phenotype distinction seen in present-day life. That is, we have our first indication of a division of labor between the part of the organism concerned with replication (in this case the RNA) and the part that interacts with the environment (the proteins).

The autocatalysis origin of life theory circumvents the 'chicken-and-egg' problem by positing that the same collective entity is both code and decoder. This entity doesn't look like a code in the traditional sense because it is a code not by design but by default. The code is embodied in the physical structures of the molecules; their shapes and charges endow them with propensities to react with or 'mutually decode' one another such that they manifest external structure, in this case a copy of its 'collective self'.

## 2.3 Autocatalytic Closure in a Cognitive System

We have considered two paradoxes—the origin of the psychological foundations of culture, and the origin of life—which from hereon will be referred to as OOC and OOL respectively. The parallels between them are intriguing. In each case we have a system composed of complex, mutually interdependent parts, and since it is not obvious how either part could have arisen without the other, it is an enigma how the system came to exist. In both cases, one of the two components is a storehouse of encoded information about a self in the context of an environment. In the OOL, DNA encodes instructions for the construction of a body that is likely to survive in an environment like that its ancestors survived. In the OOC, an internal model of the world encodes information about the self, the environment, and the relationships between them. In both cases, decoding a segment of this information storehouse generates another class of information unit that coordinates how the storehouse itself gets decoded. Decoding DNA generates proteins that orchestrate the decoding of DNA. Retrieving a memory or concept from the worldview and bringing it into awareness generates an instant of experience, a meme, which in turn determines which are the relevant portion(s) of the worldview to be retrieved to generate the *next* instant of experience. For example, if you had the thought 'my baby seems to have measles', you might rack your brain to see what you know about measles.

The bottleneck in the OOC seems to be the establishment of a network of inter-related memes, a worldview, that progressively shapes and is shaped by a stream of self-triggered thought. We want to determine how such a complex entity might come to be. Donald (1991) claims that the transition from episodic to cultural culture "would have required a fundamental change in the way the brain operates." Drawing from the OOL scenario presented above, we will posit that meme evolution begins with the emergence of a collective autocatalytic entity that acts as both code and decoder.

In the OOL case we asked: what was lying around on the primitive earth with the potential to form some sort of self-replicating system? The most promising candidate was catalytic polymers, the molecular constituents of either protein or RNA. Here we ask an analogous question: what sort of information unit does the episodic mind have at its disposal? It has memes, specifically memories of episodes. Episodic memes then constitute the 'food set' of our system.

Next we ask: what happens to the 'food set' to turn it into a self-replicating system? In the OOL case, food set molecules catalyzed reactions on each other that increased their joint complexity, eventually transforming some subset of themselves into a collective web for which there existed a catalytic pathway to the formation of each member molecule. I propose that an analogous process transforms an episodic mind into a culture-evolving one. Food set memes activate redescriptions of each other that increase their joint complexity, eventually transforming some subset of themselves into a collective web for which there exists a retrieval pathway to the formation of each member meme. Much as polymer A brings polymer B into existence by catalyzing its formation, meme A brings meme B into conscious awareness by retrieving it from memory. Note that a 'retrieval' can be reminding, a redescription of something in light of new contextual information, or a creative blend or reconstruction of many stored memes.

We know that the brains of some prehistoric tribe somehow turned into instruments for the variation, selection, and replication of memes. How might Barney, a member of this tribe, have differed from his ancestors such that he was able to initiate this kind of transformation? To answer this question, we need to briefly summon what we have learned from the neurobiology and artificial intelligence, and build a best-guess model of human cognition.

The first thing to note is that memory is *sparse*. Where $n$ is the number of features the senses can distinguish, N, the number of memes that could potentially be hosted by the focus = $2^n$ for boolean variables (and it is infinitely large for continuous variables). For example, if $n =1,000$, $N = 2^{1,000}$ memes. Since assuming $n$ is large, $N$ is enormous, so the number of locations $L$ where memes can be stored is only a small fraction of the $N$ perceivable memes. The number of different memes *actually stored* at a given time, $s,$ is constrained by $L$, as well as by the variety of perceptual experience, and the fact that meme retrieval, though distributed at the storage end, is serial at the awareness end. That is, the rate at which streams of thought reorganize the network is limited by the fact that everything is funneled through an awareness/attention mechanism; we can only figure one thing out at a time. In fact the difference between $s$ and $N$ is even greater since the mind rarely if ever attends all the stimulus dimensions it is capable of detecting.

The set of all possible n-dimensional memes a mind is capable of storing can be represented as the set of vertices (if features assume only binary values) or points (if features assume continuous values) in an *n*-dimensional hypercube, where the $s$ stored memes occupy some subset of these points. The distance between two points in this space is a measure of how dissimilar they are, referred to as the Hamming distance. Kanerva (1988) makes some astute observations about this memory space. The number of memes at Hamming distance $d$ away from any given meme is equal to the binomial coefficient of $n$ and $d$, which is well approximated by a Gaussian distribution. Thus if meme X is 111...1 and its antipode is 000...0, and we consider meme X and its antipode to be the 'poles' of the hypersphere, then approximately 68% of the other memes lie within one standard deviation (sqrt($n$)) of the 'equator' region between these two extremes (FIGURE 4). As we move through Hamming space away from the equator toward either Meme X or its antipode, the probability of encountering a meme falls off sharply by the proportion sqrt($n$)/$n$.

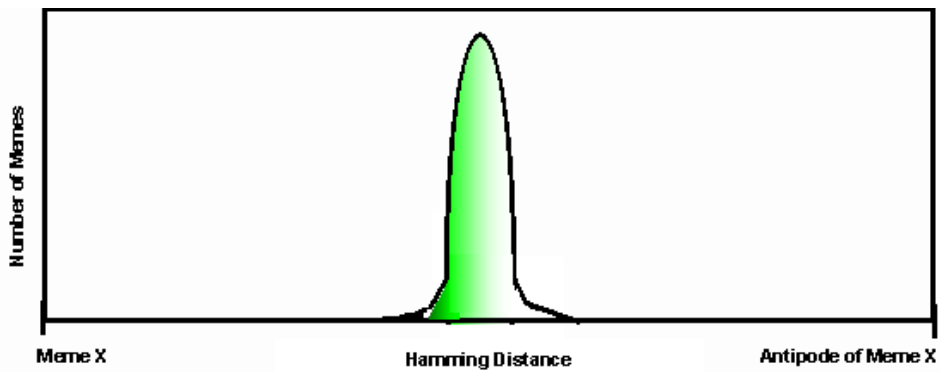

FIGURE 4. Solid black curve is a schematic distribution of the Hamming distances from address of a given meme to addresses of other memory locations in a sparse memory. The Gaussian distribution arises because there are many more ways of sharing an intermediate number of features than there are of being extremely similar or different. A computer memory stores each item in only the left-most address, whereas a distributed network stores it throughout the network. A restricted activation function, such as the radial basis function, is intermediate between these two extremes. Activation decreases with distance from the ideal address, as indicated by grey shading.

In a sparse memory, the probability that a meme one encounters is identical to one stored in memory is virtually zero. Therefore, retrieval *should* be impossible. In a neural network'a computer architecture inspired by how brains learn and retrieve information'this problem is solved by distributing the storage of a meme across many locations. Likewise, each location participates in the storage of many memes. The object of attention is represented as input/output nodes, memory locations as hidden nodes, and their pattern of connectivity as weighted links. An input touches off a pattern of activation which spreads through the network until it relaxes into a stable configuration, or achieves the desired input-output mapping using a learning algorithm. The output vector is determined through linear summation of weighted inputs. Thus a retrieved meme is not activated from a dormant state, but 'reconstructed'.

How can such a network avoid interference amongst the stored patterns? By *restricting* the distributed activation (as in a radial basis function). Recall that in the OOL case, it was crucial that the polymers be catalytic. We gave each polymer a small, random probability $P$ of catalyzing each reaction. Here we do something similar. A hypersphere of locations is activated, such that activation is maximal at the center and tapers off in all directions according to a Gaussian distribution (see FIGURE 4). The lower the *neuron activation threshold*, the wider this distribution, and therefore the more memes are activated in response to any given meme. Another way the mind prevents interference is by being *modular*; that is, different regions of the brain specialize in the processing of different kinds of information.

The final feature we will note about the brain is that it is *content-addressable*. That is, there is a correspondence between the location where a meme is stored and its semantic content. Thus each meme can only evoke, or activate, other memes that are similar to it. For example, when considering the problem of having to get out of your car every day to open the garage door, you would not think about doilies or existentialism, but concepts related to the problem—electricity, human laziness, and various openers you have encountered before.

Let us now consider what would happen if, due to some genetic mutation, Barney's activation threshold were significantly lower than average for his tribe. This means that a greater diversity of memes are activated in response to a given experience, and a larger portion of the contents of memory merge and surface to awareness in the next instant. When meme X goes fishing in memory for meme X', sooner or later this large hypersphere is bound to 'catch' a stored meme that is quite unlike X. For example, since Barney sees the sun every day, there are lots of 'sun-dominated episodes' stored in his brain. For simplicity, let us say they consist of a sequence of ten 0's followed by a five bit long variable sequence. One night he looks up into the heavens and sees the Evening Star, which gets represented in his focus as 000000011101010. This Evening Star episode will be referred to as meme X. Because the hypersphere is wide, all of the sun memories lie close enough to meme X to get evoked in the construction of X' (as is X itself). Since all the components from which X' is made begin with a string of seven zeros, there is no question that X' also begins with a string of seven zeros. These positions might code for features such as 'appears in sky', 'luminous', *et cetera*. The following set of three 1s in the 'sun' memes are canceled out by the 0s in the 'Evening Star' meme, so in X' they are represented as *s. These positions might code for features such as 'seen during the day'. The last five bits constituting the variable region are also statistically likely to cancel one another out. These code for other aspects of the experience, such as, say, the smell of food cooking or the sound of wind howling. So X' turns out to be the meme 0000000********, the generic category 'heavenly body', which then gets stored in memory in the next iteration. This evocation of 'heavenly body' by the Evening Star episode isn't much of a stream of thought, and it doesn't bring Barney much closer to an interconnected conceptual web, but it is an important milestone. It is the first time he ever derived a new meme from other memes, his first abstraction, his first creative act.

Once 'heavenly body' has been evoked and stored in memory, the locations involved habituate and become refractory (so, for instance, 'heavenly body' does not recursively evoke 'heavenly body'). However, locations storing memes that have *some* 'heavenly body' features, but that were not involved in the storage of 'heavenly body', are still active. 'Heavenly body' might activate 'moon' and then perhaps 'cloud' *et cetera*, thus strengthening associations between the abstract category and its instances. Other abstractions form in analogous fashion. As Barney accumulates both episodic memes and abstractions, the probability that any given attended meme is similar enough to some previously-stored meme to activate it increases. Therefore reminding acts increase in frequency, and eventually become streams of remindings, which get progressively longer. He is now capable of a train of thought. His memory is no longer just a waystation for coordinating stimuli with action; it is a forum for abstractive operations that emerge through the dynamics of iterative retrieval.

How do we know that streams of thought will induce a phase transition to a critical state where for some subset of memes there exists a retrieval pathway to each meme in the subset? In the OOL case, we had to show that R, the number of reactions, increased faster than N, the number of polymers. Similarly, we now want to show that some subset of the memes stored in an individual's mind inevitably reach a critical point where there is a path by which each meme in that subset can get evoked. But here, it is *not* reasonable to assume that all $N$ perceivable memes actually exist (and can therefore partake in retrieval operations). The awareness/attention filter presents a bottleneck that has no analog in the OOL scenario. As a result, whereas OOL polymers underwent a *sharp* transition to a state of autocatalytic closure, any analogous transition in inter-meme relatedness is expected to take place *gradually*. So we need to show that $R$, the diversity of ways one meme can evoke another, increases faster than not $N$ but $s$, the number of stored memes, i.e., memes that have made it through this bottleneck. That is, as the memory assimilates memes, it comes to have more ways of generating memes than the number of memes that have explicitly been stored in it.

Under what conditions does that $R$ increases faster than $s$? Once again the reader is referred to (Gabora 1998) for the mathematical details. The key idea, however, is that abstraction increases $s$ by creating a new meme, but it increases $R$ more, because the more abstract the concept, the greater the number of memes a short Hamming distance away (since irrelevant dimensions make no contribution to Hamming distance). Second, as $n$ starts to decrease the number of possible abstractions for each value of $n$ increases (up to $M/2$, after which it starts to decrease). Taken together these points mean that lower-dimensional memes enable exponentially more retrieval paths. The more deeply a mind delves into lower-dimensional abstractions, the more the distribution in FIGURE 4 rises and becomes skewed to the left. So, whereas $R$ increases as abstraction makes relationships increasingly explicit, $s$ levels off as new experiences have to be increasingly unusual in order to count as new and get stored in a new constellation of locations. Furthermore, when the carrying capacity of the memory is reached, $s$ plateaus, but $R$ does not. Thus, as long as the neuron activation threshold is large enough to permit abstraction and small enough to permit temporal continuity, the average value of $n$ decreases. Sooner or later the system is expected to reach a critical percolation threshold such that $R$ increases exponentially faster than $s$, as in FIGURE 2. The memory becomes so densely packed that any meme that comes to occupy the focus is bound to be close enough in Hamming distance to some previously-stored meme(s) to evoke it. The memory (or some portion of it) is holographic, in the sense that there is a pathway of associations from any one meme to any other. Together they form an autocatalytic set. What was once just a collection of isolated memories is now a structured network of concepts, instances, and relationships—a worldview.

Now that we have an autocatalytic network of memes, how does it self-replicate? In the OOL scenario, polymer molecules accumulate one by one until there are at least two copies of each, and their shell divides through budding to create a second replicant. In the OOC scenario, Barney shares concepts, ideas, stories, and experiences with his children and tribe members, spreading his worldview meme by meme. Categories he had to invent on his own are presented to and experienced by others much as any other episode. They are handed a shortcut to the category; they don't have to engage in abstraction to obtain it. Recall how the probability of autocatalysis in the OOL simulation could be increased by raising either the probability of catalysis or the number of polymers (since it varied exponentially with *M*). Something similar happens here. Even if Barney's son Bambam has a higher neuron activation threshold than Barney, once he has assimilated enough of Barney's abstractions, his memes become so densely packed that a version of Barney's worldview snaps into place in his mind. Bambam shares *his* worldview with his friend Pebbles, who in turn shares it with the rest of the tribe. These different hosts expose their 'copy' of Barney's original worldview to different experiences, different bodily constraints, and sculpt it into a unique internal model of the world. Small differences are amplified through positive feedback, transforming the space of viable worldview niches. Individuals whose activation threshold is too small to achieve worldview closure are at a reproductive disadvantage, and, over time, eliminated from the population. Eventually the proclivity for an ongoing stream of thought becomes so firmly entrenched that it takes devoted yogis years of meditation to even briefly arrest it.

To sum up: Kauffman's proposal that life originated with the self-organization of a set of autocatalytic polymers suggests a mechanism for how discrete memories become woven into a worldview. Much as catalysis increases the number of different polymers, which in turn increases the frequency of catalysis, reminding events increase meme density by triggering symbolic abstraction, which in turn increases the frequency of remindings. And just as catalytic polymers undergo a phase transition to a state where there is a catalytic pathway to each polymer present, and together they constitute a self-replicating set, memes undergo a phase transition to a state where each meme is retrievable through a pathway of remindings/associations. Together the memes now constitute a transmittable worldview, an internalized tapestry of reality, that both weaves, and is woven by, threads of experience.

## 3. MENDING, PATCHING, AND REWEAVING

### 3.1 Abstractions Both Create and Alleviate Worldview Inconsistencies

It would seem that in the transition from the episodic mind to the culture evolving mind, we have made enormous progress. And in fact a worldview is enormously useful; with it we can imagine and evaluate the outcomes of various actions, compose music, and write articles about worldviews. It is tempting to think that our understanding of the world must be enormously more accurate than that of any other species. In fact, however, the episodic mind has *no* inconsistencies. Since it doesn't represent relationships, it doesn't get any relationships wrong. It never encoded the sun and the stars as different kinds of entities in the first place, so if it were to go further and further away from the sun until it looked just like any other star, no conceptual adjustment would have to be made. Thus there is a tradeoff between abstraction and accuracy.

This tradeoff affects the accuracy of not only impersonal beliefs, but also items of a more personal nature. Although most papers in this volume deal with inconsistencies that take the form of logical fallacies, inconsistencies in the realm of self-perception and social cognition probably exert as large an effect. For example, ever since the concepts 'good' and 'bad' came into being, many people have simultaneously held (to some extent at least) both the beliefs 'I am good' and 'I am bad'.

As the horizon of a worldview push forward, and it gets pulled to cover more and more ground, inevitably holes appear that need to be mended. Sometimes just stretching an existing concept does the trick. For example, although sun and star were originally classified as different kinds of objects, they eventually came to be classified as instances of the same thing. Or sometimes the worldview needs a metaphor. An abstraction in another part of the 'reality fabric' is called in to patch things up, as in 'The moon was a ghostly galleon'. Often the metaphor sticks, as in 'sunny disposition' or 'star pupil'. In still other cases, an abstraction is newly-invented to do the job. For example, the concept 'heavenly body' was coined so that one might refer to any of the various objects that appear in the sky.

### 3.2 Thought Trajectories and Attractors in a Cultural Fitness Landscape

Inconsistencies do not just manifest at random here and there in conceptual space. Since the culture-evolving system grew out of, and is dependent on, the biology-evolving system, the most stable relationship between them is one of mutually-beneficial coevolution. Since many, if not all, of our needs either have a biological basis (e.g., the need for food, for shelter, *et cetera*) or are derived from other needs that do (e.g., the need to make money), the generation of new memes is largely constrained by our heritage as products of biological evolution. Cultural trajectories do sometimes diverge from the 'safest' regions of conceptual space (e.g., the decision to commit suicide). These situations are unstable, caused by forces that work in opposition to the pull of the attractor. Thus another pitfall of having a worldview is that it occasionally leads its bearer astray from the actions that would most benefit it. You don't need an interwoven worldview to seek food when you are hungry and to hide when you sense danger; stimulus-response associations do the job. Thus, much as a physical organism sometimes falls down when the impetus that sets a limb into motion ignores contextual input from the rest of the body, a worldview runs the risk of losing its 'conceptual center of balance' when the thought trajectories it spawns stray far from basic human concerns. So worldviews tend to grow (and thus inconsistencies manifest) in regions of conceptual space that deal with human affairs.

This brings us to the notion of a *fitness landscape*. Consider the vast space of possible memes discussed in Section 3. Some of them are more useful for a human than others. A fitness landscape is a measure of the relative value, or *fitness*, of each possibility given some set of constraints. Fitness landscapes are portrayed graphically by choosing two important dimensions and plotting them on the X and Z axes, and plotting fitness on the Y axis. Since points that have similar values for dimensions X and Z tend to have similar fitnesses (in other words, the landscape is correlated), fitness landscapes look like pieces of fabric, with rolling hills and valleys. Cultural landscapes are in a state of perpetual flux as one need gets satisfied and another takes precedence.

We can think of a worldview as a region of conceptual space where, given the needs we routinely experience, our cultural trajectories tend to dwell. Our worldviews overlap to the extent that similar genetic heritage and experience invite similar cultural trajectories. But each person's temperament and experience are different, so each person's private worldview carves out a unique path through conceptual space. This is obvious when we engage in the subtle interplay of cultural exchange; everyone has a different personality, leaves us with a different 'flavor'. One can think of a worldview as a personality as it is experienced from the inside.

Abstractions are not only driven by the cultural fitness landscape, they feed back on and actually alter its topology. Much as the evolution of rabbits created ecological niches for species that eat them and parasitize them, the invention of cars created cultural niches for gas stations, seat belts, and garage door openers. As one progresses from infant-hood to maturity, and simple needs give way to increasingly complex needs, the trajectory of a stream of thought acquires the properties of a chaotic or strange attractor. The landscape is fractal (i.e., there is statistical similarity under change of scale) in that the satisfaction of one need creates other needs. This is analogous to the fractal distributions of species and vegetation patterns described by ecologists (Mandelbrot (60), Palmer (69), Scheuring & Riedi (83)).

## 3.3 Multiple Layers of Abstraction

In the OOL case, since short, simple molecules are more abundant and readily-formed than long, complex ones, it made sense to expect that the food set molecules were the shortest and simplest members of the autocatalytic set that eventually formed. Accordingly, in simulations of this process, the 'direction' of novelty generation is outward, joining less complex molecules to form more complex ones through AND operations (see FIGURE 5). In contrast, the elements of the cognitive "food set" are complex, consisting of all attended features of an episode. In order for them to form an interconnected web, their interactions tend to move in the opposite direction, starting with relatively complex memes and forming simpler, more abstract ones through OR operations. The net effect of the two is the same: a network emerges, and joint complexity increases. But what this means for the OOC is that there are numerous levels of autocatalytic closure, which convey varying degrees of worldview interconnectedness and consistency on their 'meme hosts'. These levels correspond to increased penetration of the ($n$-1, $n$-2')-dimensional nested hypercubes implicit in an $n$-dimensional memory space. Since it is difficult to visualize the set of nested, multidimensional hypercubes, we will represent this structure as a set of concentric circles, such the outer skin of this onion-like structure represents the hypercube with all $n$ dimensions, and deeper circles represent lower-dimensional hypercubes (FIGURE 5). Obviously, not all the nested hypercubes can be shown. The points of our original hypercube are represented as points along the perimeter of these circles, and the centermost location where a meme is stored is shown as a large, black dot.

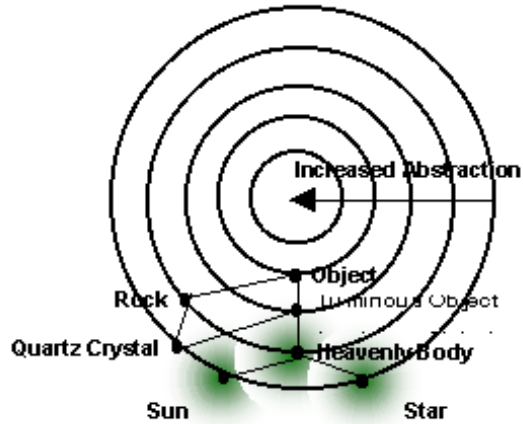

FIGURE 5. The role of abstractions in creative thought. For ease of visualization, the set of nested hypercubes representing the space of possible memes is shown as a set of concentric circles, where deeper circles store deeper layers of abstraction (lower dimensional hypercubes). A black dot represents the centermost storage location for a specific meme. 'Heavenly body' is a more general concept than 'sun' or 'star', and is therefore stored at a deeper layer of abstraction. Green circle around each stored meme represents hypersphere where the meme gets stored and from which the next meme is retrieved.

The outermost shell encodes memes in whatever form they are in the first time they are consciously encountered. This is all the episodic mind has to work with. In order for one meme in this shell to evoke another, they have to be extremely similar at a superficial level. In a cultural mind, however, related concepts are within reach of one another because they are stored in overlapping hyperspheres.

'Sun' and 'star' might be too far apart in Hamming distance for one to evoke the other directly. However, by attending the abstraction 'Heavenly Body', which ignores the 'seen at night versus seen during the day' distinction, the mind decreases the apparent Hamming distance between them.

The most primitive level of autocatalytic closure is achieved when stored episodes are interconnected by way of abstractions just a few 'onionskin layers' deep, and streams of thought zigzag between these superficial layers. A second level occurs when relationships amongst *these* abstractions are identified by higher-order abstractions at deeper onionskin layers. *Et cetera*. Once an individual has defined an abstraction, identified its instances, and chunked them together in memory, she can manipulate the abstraction much as she would a concrete episode.

Undoubtedly there is selective pressure for parents who monitor their child's progress in abstraction and interact with the child in ways that promote the formation of new abstractions the next level up. Recall the discussion in Section 2 concerning the incorporation of new polymer species by supracritical autocatalytic sets. This kind of parental guidance is analogous to handcrafting new polymers to be readily integrated into a particular autocatalytic set; in effect it keeps the child's mind perpetually poised at a supracritical state.

## 3. 4 Constraint Satisfaction and Cognitive Annealing

Sometimes a new concept fits readily into the existing worldview and single-handedly resolves an inconsistency. Other times, the solution is not so simple. The canonical example is the pre-Copernican view of the universe; one discrepancy after another reared its ugly head. In situations like this, priorities must be re-evaluated, and abstractions that have long served us well must be discarded. An extensive portion of the worldview must be ripped apart and put back together in a new way.

The development of a worldview that accurately portrays reality and effectively navigates us through the maze of human affairs is a good example of what computer scientists refer to as a *multi-objective optimization* or *constraint satisfaction* problem. In other words, it involves simultaneously maximizing or minimizing multiple frequently-conflicting criteria. This kind of problem is particularly difficult to solve because the trade-offs between the multiple objectives or constraints are often unknown. The best that can be done is to find a *Pareto optimal solution*'one that cannot be improved with respect to any one objective without worsening some other objective (Steuer, 1986). A compromise is reached.

*Simulated annealing* is a mathematical technique for solving this kind of problem. It was inspired by the physical process of annealing, during which a metal is heated and then gradually cooled. The shapes and charges of the atoms confer on them varying degrees of attraction and repulsion toward one another. Configurations that maximize attractive forces and minimize repulsive ones are most stable; thus, the shapes and charges of the atoms define the constraints of the system. Heating decreases the stability of the forces that bind the atoms together'it loosens global structure'whereas cooling has the opposite effect. The slower the cooling process, the more opportunity the atoms have to settle into a stable, low-energy arrangement.

Simulated annealing computationally mimics the cooling process by decreasing and then gradually increasing the stability of the connections amongst the parts of a system in a series of either random or deterministic updates. If there are too few update steps, the system settles on a state wherein few of the constraints imposed by the structure and dynamics of its components are met. It may, for example, result in islands of mutually compatible components which are themselves incompatible, in which case the system has difficulty functioning as a whole. The greater the number of updates, the more harmonious the state the system eventually settles into, i.e., the more likely it is to find a Pareto optimal solution. An interesting feature of this proces is that as the temperature is lowered, the *correlation length*'that is, how far apart the components of a system must be before their mutual information falls to zero'increases. Another way to say this is: the *mutual information*'the amount of information that can be gleaned about one component of a system by examining another component'increases. The result is that a perturbation to any one component can percolate through the system and affect even distant components.

How does this pertain to worldviews? When inconsistencies spring up all over the place, simply assimilating a new concept or two is insufficient. Large-scale worldview renovation is in order. It seems reasonable to expect that when this happens there would be a tendency to temporarily 'loosen' one's internal model of reality, weaken inter-meme relationships, so as to allow new insights to more readily percolate through and exert the needed revolutionary impact. Then one slowly 'anneals' as the details of how to best structure this new and improved worldview fall into place. This could be achieved by decreasing the neuron activation threshold'thus increasing the potential of any meme to trigger a chain reaction of novel associations'and then slowly increasing the threshold, thereby stabilizing associations that are consistent and fruitful. In fact, simulated annealing is sometimes used to improve the performance of a neural network (e.g., Cohen 1994; Cozzio 1995).

What sort of meme might be able to capitalize on this state of increased readiness, and trigger cognitive reorganization of this magnitude? Strong candidates include the logical operators 'and', 'or', and 'not'. When a child first glimpses how these abstractions can be used to manipulate and invent symbols, the potential for dramatic reconfiguration and overall expansion of the conceptual network may well be greater than at any other time in his or her life. The assimilation of other particularly useful abstractions, such as 'mine', 'depth', or 'time', as well as frames (Barsalou, in press), scripts (Schank & Abelson 1977), and schemas (Bartlett 1932, 1958; Piaget 1926, 1936/52; Minsky 1985) might also be expected to precipitate substantial worldview revamping.

The phenomenon need not be limited to concepts we acquire as a child (although it seems reasonable to expect that the resolution and fidelity with which our worldview has been formed each step of the way constrains the number and magnitude of revisions it will later undergo). Science offers abundant potential for this. The more inconsistent an experimental result, the greater its potential to induce such a transformation. The fact that quantum mechanics defies complete integration into our worldview, for example, suggests that its transformation potential is great.

Concepts that have the capacity to exert a dramatic effect are probably rare. Rosch's (1978) work on basic level categories suggests that the way we develop abstractions and use them to organize information is not arbitrary but emerges in such a way as to maximize explanatory power. It would not be surprising to find that the relationship between concept frequency and degree of abstraction exhibits the same kind of power law relationship as one finds in other emergent systems (Bak, Tang, & Weisenfeld 1988).

## 4. FLAWS AND FOLDS

### 4.1 How Censorship Fragments the Worldview

The process of conceptual re-annealing described above is expected to take place as a last resort only. Why? Because it makes the worldview vulnerable. It allows an alien element to deeply penetrate a workable system, with unforseen and potentially harmful consequences. Just as importing foreign plants can bring ecological disaster, assimilating a foreign meme can upset the established state of harmony in a conceptual network, thereby inviting confusion and depression.

In fact, a child develops *mental censors* that ward off the internalization of potentially threatening memes (Minsky1985). This includes memes that have the potential to disrupt the belief structure (such as the idea of natural selection to a creationist) as well as memes that could damage the ego (such as the realization that you are ugly). It includes any meme that could in some way threaten survival (such as the realization that you don't believe in the product your company produces, and a host of other memes that I can't tell you about because my censors prohibit them).

Censorship could easily be accomplished through a procedure opposite to that described in Section 3.3: temporarily *increasing* the activation threshold, and thereby prematurely terminating assimilation of the current contents of awareness into the conceptual network. The censored meme would then be isolated from the memory at large, much as are the episodic memes in a primate's brain. Censoring a meme alters both the probability that it gets evoked (activated into awareness) by other memes, and the probabilities involved in determining which other memes are evoked by *it* once it has become active. It is like a portion of fabric that is fenced in on all sides by knots. Pulling on a fiber of the fabric *outside* the fenced region does not exert much of a pull on that fiber *inside* the fenced region, and vice versa; the knots dampen the force of the pull by diffusing it across the tangled mass of other fibers. Much as erecting a real fence increases the probability that people will stay on either one side or the other, the censored meme is either avoided, or dwelt on excessively. This is consistent with our bipolar attitude toward highly censored subjects such as aggression and sexuality, and seems to correspond closely to what psychiatrists refer to as altered schema valence, wherein specific topics elicit in the patient either *latent valency* (excessive avoidance), or *hypervalency* (excessive preoccupation) (Beck & Freeman 1990).

Extensive censoring is referred to as *repression*. Recall our discussion of simulated annealing, where we introduced the term correlation length. If the above line of reasoning is correct, repression lowers the mutual information and the average correlation length between memes. The individual is less able to respond spontaneously because contextual information takes longer to percolate through the network. This then sheds some light on how self-destructive memes might emerge. Much as it would be hard to stay physically balanced if the nerve endings from one of your legs were blocked, it is hard to stay psychologically balanced if significant portions of your conceptual network are fenced off. It also provides us with a relatively tangible interpretation of the common phrases 'psychologically unstable', 'fragmented reality' and 'split personality'.

These processes appear to take place not only at the level of an individual's worldview, but also at the level of the collective worldview of a society. Once again we have a trade-off. This time it is between social mores that encourage free thinking and thereby risk the proliferation of potentially dangerous thought trajectories, and social mores that discourage free thinking, at the risk of increased repression and deception.

### 4.2 How Deception Invites Worldview Distortion

We have all felt at one time or another the strain of telling a lie, or of living a lie, as in by pretending to like someone we do not like. Once you lie, you never know what other 'threads of reality' will have to be drawn in to maintain consistency. For example, the cheating spouse says she was working late at the office. Her husband asks why she didn't answer her office phone. She lies again and says that she was in her colleague's office because her computer is broken. And so on.

Whereas repression halts the assimilation process, deception blocks thoughts that have already been assimilated. Much as a fold in one part of a piece of fabric induces folding in adjacent regions, deception perpetuates more deception in nearby regions of conceptual space. The folded-over, hidden-from-view portion of the worldview is avoided, much as censored material is, with the ensuing fragmentation and loss of conceptual balance described above.

Deception is widespread in not only humans (Mitchell & Thompson 1986), but also plants and animals (e.g., Dawkins 1982; Krebs & Dawkins 1984; Wallace 1973). By misleading others, the perpetrator of deception gains an unfair advantage. Some have argued that not only deception, but *self*-deception is adaptive, because it enables one to more convincingly deceive others (Alexander 1975, 1979; Trivers 1976, 1985). As Trivers puts it, "there must be strong selection for a degree of self-deception, rendering some facts and motives so as not to betray—by the subtle signs of self-knowledge—the deception being practiced." He goes on to argue that if deception increases fitness, and self-deception increases ability to deceive, then: "the conventional view that natural selection favors nervous systems which produce ever more accurate images of the world must be a very naïve view of mental evolution."

One might conclude that deception is an unavoidable feature of the human condition. When cognition is viewed as a constraint satisfaction problem, we see that this isn't necessarily the case. Each individual's worldview can be seen as a different solution to the often-conflicting demands of human survival. Our worldviews are constantly evolving as different approaches to life play themselves out in our various endeavors and interactions. Of course I may be deceiving myself, but it seems to me that if the ideas presented above hold any merit, the final word is not yet in as to whether deception will play a part in the most optimal solutions humanity eventually zeros in on.

## 5. REFLECTIONS ON THE CONCEPTUAL TAPESTRY

There is a web called Indra's Net, made of threads of light. It stretches horizontally through space, and vertically through time. At every intersection dwells an individual, and at every individual lies a crystal bead of light.

--Buddhist allegory

We have been developing a metaphor between the properties of an interconnected worldview and the properties of a piece of woven fabric. The metaphor has served to sharpen our intuitions about less tangible cognitive phenomena. From what sort of fibers might the conceptual web be woven from?

One unusual property of memes is that they can readily and instantaneously combine with one another to create something new. Another unusual property of memes is their dual nature. The fabric of reality is continuously stretched, pulled back on itself, never the same. It can viewed as a set of interconnected memes, a perspective that focuses on the individual units of which it is composed. Alternatively it can be viewed as an emergent whole, a perspective that focuses on the undulations that run across its length as one thought after another rises to the foreground and falls back giving way to another.

There is another substance that exhibits these properties, one that in fact has been used as an analog of cognition since the dawn of civilization, and that is light. The metaphor permeates our language still, as in: moment of illumination, he beamed, her face lit up, to glow with enthusiasm, flash of insight, ray of hope, dim-whitted person, light of my life, show me the light, *et cetera*. Let us then briefly explore whether this analogy can provide insight into the ideas we have been developing.

## 5.1 Bending the Truth

We will begin by addressing the issue of deception discussed in Section 4.2. When a beam of light passes from one medium to another perpendicular to the boundary between them, as in FIGURE 6a below, it passes straight through without refracting (bending). When the beam of light passes from one medium to another at any *other* angle, as in the figure below, it refracts. The greater the deviation from 90', or the greater the difference in density between the two media, the greater the refraction (FIGURE 6b).

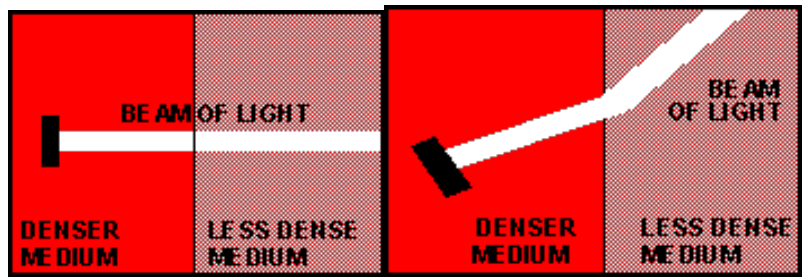

Figure 6a. Light passing from one medium to another at an angle perpendicular to their boundary does not refract. 6b. Light passing from one medium to another at an angle that is *not* perpendicular to their boundary *does* refract.

The light metaphor offers a speculative suggestion for why the proclivity to deceive others is highly correlated with a distorted perception of reality (as in neuroses) (Beck et al., 1990). If in FIGURE 6b above, the light from the less dense medium were projected *back* to the dense media, on this inward trip there would once again be refraction. Analogously, if one gets into the habit of orienting one's thought trajectories such that thoughts come *out* distorted (for example, in order to more acceptable to others) then perhaps, if that orientation becomes habitual, things start coming *in* distorted.

## 5.2 Focusing Attention and Reflecting on an Idea

In fact, distortion is not confined to the situation wherein someone tells a lie; a more subtle form is inherent in *all* communication. The act of transforming a raw idea into words, gestures, notes, or even equations, necessarily distorts it. There is no way to avoid this kind of distortion, short of never expressing ideas.

Finding a way of expressing a meme that minimizes distortion while keeping the meaning intact takes time and effort.

Once again, light exhibits an analogous behavior. When a wide beam passes through a concave surface, it becomes more diffuse, as in FIGURE 7a. This is because no matter what direction the beam is pointed, there is no way it can meet the surface at a perpendicular angle across its full width. Therefore the edges are forced to bend, or refract.

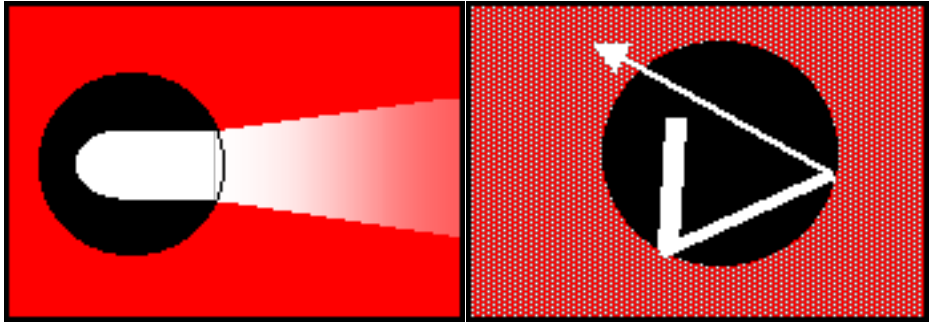

Figure 7. A conceptual metaphor for (a) the distortion and decrease in intensity that takes place when a vague idea is expressed, and (b) the clarification and heightened intensity that occurs if the idea is first reflected upon.

When a beam of light strikes a boundary between two media, it does not always pass through. Sometimes it *reflects* off the boundary surface back into the first media. When a diffuse beam of light is repeatedly reflected off a concave surface such as the interior of a sphere, it becomes more focused, as in FIGURE 7b. Similarly, when we reflect on an initially vague idea, it becomes more focused in our minds. Reflecting on an idea amounts to reflecting it back and forth off 'onionskin layers' of varying degrees of abstraction, refining it in the context of its various interpretations.

## 5.3 Superficial and Deep Ideas

Are all ideas equally prone to this kind of distortion? Let us return to the 'layers of abstraction' concept discussed in Section 3.3. Deep ideas, like the concept of *opposite,* or of *depth* itself, apply to almost any domain of life. Similarly, light that originates in the center of a sphere can radiate outward in any direction without refraction, because wherever it contacts the sphere it is perpendicular to it, as in FIGURE 8a.

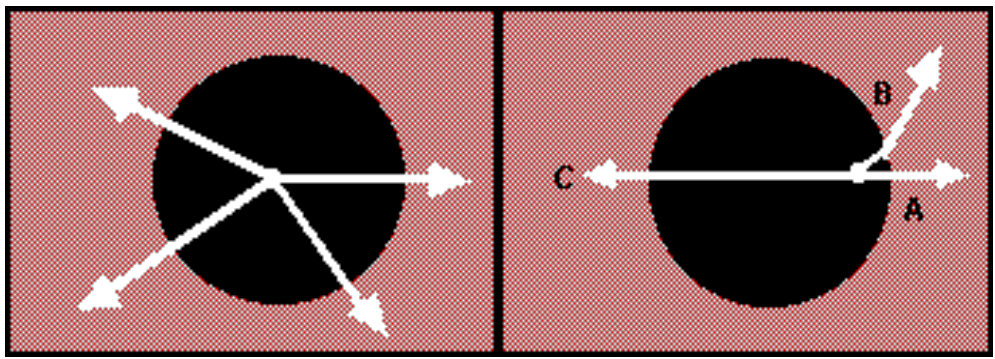

Figure 8. A conceptual metaphor for (a) the ubiquitous applications for deep concepts versus (b) the domain-specific nature of superficial ones.

Superficial ideas are relevant to only a narrow domain of life; they don't translate well to some other subject matter. Once again we invoke a light metaphor. Light that originates near the surface of a sphere can only radiate through a narrow region without reflection and refraction. Thus in FIGURE 8a, when light hits the sphere at a perpendicular angle, as in Beam A, there is no refraction. If the angle is just a little bit off, as in Beam B, it refracts. Strangely enough, if it travels *backwards* as in Beam C (in the opposite direction to Beam A), there is once again no refraction. This concurs with the longstanding yet counterintuitive notion of unity in opposites. The fact that we are reaching a point in scientific history where interdisciplinary research is commonplace, and concepts that apply to all the sciences'such as chaos, fractal, attractor, and fitness landscape'are almost cliché (and quickly working their way into the layperson's vocabulary), may be an indication that the scientific component of our worldview is approaching that innermost core of conceptual abstraction. (Of course if we code episodes at a higher resolution to begin with, even deeper levels of abstraction may await us')

It may be that these metaphors are useful only as conceptual tools. However it may be that the mathematical basis of the two phenomena may have much in common. If this is the case, we can expect them to provide one another with 'conceptual scaffolding', as do the concepts of biological and cultural evolution. Work is underway to investigate how deeply the correspondence goes. Since light has been used as a metaphor for cognition since the dawn of civilization, it would be very interesting these systems are found to be, at an abstract level of analysis, the same thing.

## 6. CONCLUSIONS

We have looked at a tentative model to explain how an interconnected conceptual web, or worldview, could emerge from a collection of discrete episodic memories through the formation and progressive elaboration of abstractions, and the identification of relationships. We have seen that there is a cost to this transformation; without a worldview there are no inconsistencies. Inconsistencies can often be mended, or patched up with abstractions. But there is no guarantee that the human mind is capable of weaving its memories and abstractions into an entirely consistent construction, particularly given our tendencies toward repression and deception. Perhaps the 'enlightened' thing to shoot for is not a worldview that is particularly extensive, or even completely accurate, but one which unfolds in its own time while allowing us to maintain our conceptual center of balance.

## ACKNOWLEDGEMENTS

I would like to thank Diederik Aerts, Didier Durlinger, and William Macready for helpful comments and discussion. I would also like to acknowledge the support of the Center Leo Apostel.